\documentclass[12pt]{iopart}
\usepackage{graphicx}
\usepackage{epsfig}
\usepackage{psfrag}
\usepackage[usenames]{color}
\usepackage{xcolor}
\usepackage[colorlinks=true,linkcolor=blue,citecolor=blue]{hyperref}
\usepackage{bm}
\usepackage{bbm}
\usepackage{appendix}
\expandafter\let\csname equation*\endcsname=\relax
\expandafter\let\csname endequation*\endcsname=\relax
\usepackage{amsmath}
\usepackage{physics}

\begin{document}

\title{Bogolon-mediated light absorption in atomic condensates of different dimensionality}

\author{D.~Ko$^{1,2}$, M.~Sun$^{3,1}$, V.~M.~Kovalev$^{4,5}$, I.~G.~Savenko$^{1,2}$}
\address{
{\small $^1$ Center for Theoretical Physics of Complex Systems, Institute for Basic Science (IBS), Daejeon 34126, Korea} \\
{\small $^2$ Basic Science Program, Korea University of Science and Technology (UST), Daejeon 34113, Korea}\\
{\small $^3$ Faculty of Science, Beijing University of Technology, Beijing 100124, China.}\\
{\small $^4$ A.V. Rzhanov Institute of Semiconductor Physics, Siberian Branch of Russian Academy of Sciences, Novosibirsk 630090, Russia}\\
{\small $^5$ Novosibirsk State Technical University, Novosibirsk 630073, Russia}
}

\date{\today}

\begin{abstract}
In the case of 
structureless bosons, cooled down to low temperatures, the absorption of electromagnetic waves by their Bose-Einstein condensate is usually forbidden due to the momentum and energy conservation laws: the phase velocity of the collective modes of the condensate called bogolons is sufficiently lower than the speed of light.
Thus, only the light scattering processes persist.
However, the situation might be different in the case of composite bosons or the bosons with an internal structure.
Here, we develop a microscopic theory of electromagnetic power absorption by a Bose-Einstein condensates of cold atoms in various dimensions, utilizing the Bogoliubov model of a weakly-interacting Bose gas. 
Thus, we address the transitions between a collective coherent state of bosons and the discrete energy levels corresponding to excited internal degrees of freedom of non-condensed individual bosons. It is shown, that such transitions are mediated by one and two-bogolon excitations above the condensate, which demonstrate different efficiency at different frequencies and strongly depend on the condensate density, which influence varies depending on the dimensionality of the system.
\end{abstract}

\section{Introduction}

Radiation pressure generates a flow of particles, for instance, electrons, molecules or atoms due to the momentum transfer from the photons into the system.
The direction of the resultant, generated by the light field current of particles coincides (up to the sign) with the direction of the  wave vector of the light field, $\textbf{j}(\omega)\sim \textbf{k}\alpha(\omega)I$, where $\textbf{k}$ is the photon wave vector, $I$ is the intensity of the electromagnetic (EM) wave and $\alpha(\omega)$  is the absorption coefficient with $\omega$ the EM field frequency. 
This phenomenon is often referred to as the photon drag effect.
It has been widely studied in generic two-dimensional (2D) electron gas and graphene systems~\cite{PhysRevLett.64.463, GLAZOV2014101}, metal films~\cite{PhysRevLett.123.053903,Khichar_2021}, topological insulators~\cite{LEE201644, Plank2016PhotonDE}, van der Waals structures and Dirac materials~\cite{KiemleZimmermannHolleitnerKastl+2020+2693+2708, PhysRevB.103.L161301}, and cavities~\cite{PhysRevB.82.125307}.

In cold atoms, the role of radiation pressure-based techniques cannot be overestimated.
The primary reason why the radiation pressure is so important there is the possibility of particle manipulation and trapping by external fields~\cite{PhysRevLett.59.2631, PhysRevLett.40.729}, resulting in their confinement and focusing or acceleration~\cite{PhysRevLett.55.48, PhysRevLett.41.1361, PhysRevLett.24.156, PhysRevLett.68.1492, PhysRevLett.78.4709}.
This allows for, in particular, effective laser cooling~\cite{PhysRevLett.61.169, PhysRevLett.74.3348, RefHarrisNatPh, RevModPhys.70.707} by utilizing several coherent laser beams with the possibility of subsequent formation of an atomic Bose-Einstein condensate (BEC)~\cite{doi:10.1126/science.269.5221.198}.

In the normal state of an atomic gas, it experiences two fundamental processes of light-matter interaction, i.e. light scattering and light absorption.  
However, once the BEC is formed, the interaction of light by the system gets suppressed. 
Indeed, according to, e.g., the Bogoliubov model of a weakly interacting Bose gas, since the over-BEC excitations of condensate density fluctuations, bogolons, possess linear dispersion with their velocity much smaller than the speed of light, the system absorption violate the conservation laws.  
That is why only the light scattering processes have been actively studied in cold atom BECs. 

However, if we consider Bose particles with  internal degrees of freedom, then finite absorption can happen involving the transitions between the energy states of internal particle motion.
Then, the system still experiences the influence of external light field after the formation of the BEC.
Indeed, the spectrum of a single atom with an eigenfunction $|\eta,\textbf{p}\rangle$ reads
$E_\eta(\textbf{p})=E(\textbf{p})+\Delta_\eta$, where
$E(\textbf{p})=\textbf{p}^2/2m$ 
is a kinetic energy of the particle center-of-mass motion, and $\Delta_\eta$ is the energy spectrum of the internal motion of the atom; 
the index $\eta$ stands for the full set of quantum numbers  characterizing the internal spectrum of the particle, thus, the value $\eta=0$ refers to the ground state of the internal spectrum of Bose particles. 

%All the particles are in the ground $\eta=0$ state of their internal motion, zero kinetic energy $\textbf{p}=0$ of their center-of-mass motion without a dipole moment.

In our subsequent analysis, we assume that initially, the system is in the BEC state, $|\eta=0,\textbf{p}=0\rangle$, and study a particular type of the radiation pressure phenomena to occur in atomic condensates of different dimensionalties.
To quantify the effect, we calculate the absorption probability and the absorption  coefficient of the system and analyse their behavior.
Obviously, the frequency dependence of the radiation pressure is determined by the spectrum of the absorption coefficient. 
This dependence is usually either monotonous or resonant: If the external EM field frequency $\omega$ approaches the energy of a quantum transition in the system, the response might experience resonant behavior.

\section{System Hamiltonian and light--bosons interaction}
%In our theoretical model, without losing any generality, we consider the bosonic field operator as a composition $\psi_\eta(\mathbf{r},t)\chi_\eta$.
%In this presentation, the center-of-mass motion of the Bosonic particle is $\psi_\eta(\mathbf{r},t)$ and internal particle motion ($\chi_\eta$) fields, where $\eta$ is a quantum number describing an internal degree of freedom. 
%In our theoretical model, without losing any generality, 
Let us consider a bosonic field operator as a composition of the center of mass term, $\psi_\eta(\mathbf{r},t)$, and an internal motion term, $\chi_\eta$, respectively. 
Thus, the full operator for the bosonic particle is $\psi_\eta(\mathbf{r},t)\chi_\eta$, where $\mathbf{r}$ is the center of mass coordinate and $\eta$ is the quantum number to represent the particle's internal degrees of freedom.
Then, the  total Hamiltonian of the system reads~\cite{PhysRevB.98.165405},

% %%%%%%%%%%%%%%%%%%%%%%%%%%%%%%%%%%%%%%%%%%%%%%%%%%%%%%%%%%%%%%%%%%%%
\begin{equation}
\label{EqH}
\hat{H}=\hat{H}_0+\hat{V},     
\end{equation}
where
\begin{equation}
\label{EqV}
\hat{V}=-\hat{\mathbf{d}}\cdot \hat{\mathbf{E}}
= -\sum_{\eta\eta^\prime} \mathbf{d}_{\eta^\prime\eta} \int d\mathbf{r}\psi_{\eta^\prime}^\dagger(\mathbf{r},t) \hat{\mathbf{E}}(\mathbf{r},t) \psi_\eta(\mathbf{r},t)    
\end{equation}
describes interaction between the bosons and light. In Eq.~\eqref{EqV}, $\hat{\mathbf{E}}(\mathbf{r},t)=\hat{\mathbf{E}}_0\exp{(\text{i}\mathbf{k}\cdot\mathbf{r}-\text{i}\omega_{\mathbf{k}}t)}+\textrm{c.c.}$ according to the classical representation of the light field with $\omega_\mathbf{k}=c|\mathbf{k}|$, and, the matrix elements of the dipole moment operator read as $\mathbf{d}_{\eta^\prime\eta}=\langle \chi_{\eta^\prime}|\hat{\mathbf{d}}|\chi_{\eta} \rangle$.

Integrating out the internal particle motion variables yields bare Hamiltonian of the system,
% %
% %
% %
% %
% %

\begin{gather}
\label{EqH0}
\hat{H}_0=\int d\mathbf{r}\psi_0^\dagger(\mathbf{r},t)
\Bigl[
\frac{\hat{{\bf p}}^2}{2m}-\mu+g|\psi_0(\mathbf{r},t)|^2
\Bigr]
\psi_0(\mathbf{r},t)\\
\nonumber
+\sum_{\eta\neq 0}\int d\mathbf{r}\psi_{\eta}^{\dagger}(\mathbf{r},t)
\Bigl[
\frac{\hat{{\bf p}}^2}{2m}+\Delta_\eta
\Bigr]
\psi_\eta(\mathbf{r},t),
\end{gather}
% %
% %
% %
% %
% %
where $\mu$ is the chemical potential and $g$ is the particle-particle interaction strength for the ground state of the bosonic system;
%and $\Delta_\eta$ is .
for simplicity and clarity of results, let us assume that the spectrum of the internal motion of particles is equidistant, $\Delta_\eta=\eta\Delta$.
The employment of a realistic spectrum of particular atoms, such as hydrogen-like spectrum, $\Delta_\eta=\Delta/\eta^2$, is a trivial complication.

The first line in Eq.~\eqref{EqH0} represents the Gross-Pitaevskii (GP) equation describing atomic BEC in the ground state with $\eta=0$ and ${\bf p}=0$, and the second line describes the bosons in the excited internal states $\eta\neq 0$. 
We will assume that most of the particles are in the condensate with the quantum number $\eta=0$, thus, the interaction of non-condensed particles with $\eta\neq 0$ with each other and with the BEC is weak and can be disregarded. 

The system described by the Hamiltonian~\eqref{EqH0} possesses two types of  low-energy excitations. 
The BEC described by the GP equation is characterized by sound-like excitations of its density (bogolons) and single-particle excitations describing the motion of individual atoms with $\eta\neq0$, as indicated by the second term in~\eqref{EqH0}. 
% The field operator also contains two terms, $\psi_0$ and $\psi_{\eta\neq0}$, corresponding to these two types of excitations in the system, respectively. 
% We assume that $\abs{\psi_0}^2\gg\abs{\psi_{\eta\neq0}}^2$, thus indicating that at low temperatures most of particles are in the BEC state. 
The field operator also contains two terms, $\psi_0$ and $\psi_{\eta\neq0}$, corresponding to these two types of excitations in the system, respectively. 
We assume that $|\psi_0|^2\gg|\psi_{\eta\neq0}|^2$, thus indicating that at low temperatures most of particles are in the BEC state. 

For legibility, it is convenient to separate the terms containing $\eta= 0$ and $\eta\neq 0$ in Eq.~\eqref{EqV},

\begin{align}
\label{EqTerms}
\hat{V}&=\hat{V}_1+\hat{V}_2+\hat{V}_3+\hat{V}_4\\ \nonumber
&= -\mathbf{d}_{00}
\int d\mathbf{r}\psi_{0}^\dagger(\mathbf{r},t) \hat{\mathbf{E}}(\mathbf{r},t) \psi_0(\mathbf{r},t)-\sum_{\eta\neq 0} \mathbf{d}_{0\eta} \int d\mathbf{r}\psi_{0}^\dagger(\mathbf{r},t) \hat{\mathbf{E}}(\mathbf{r},t)
\psi_\eta(\mathbf{r},t)\\ \nonumber
&-\sum_{\eta^\prime\neq 0} \mathbf{d}_{\eta^\prime 0} \int d\mathbf{r}\psi_{\eta^\prime}^\dagger(\mathbf{r},t) \hat{\mathbf{E}}(\mathbf{r},t) \psi_0(\mathbf{r},t)-\sum_{\eta,\eta^\prime\neq 0} \mathbf{d}_{\eta^\prime \eta} \int d\mathbf{r}\psi_{\eta^\prime}^\dagger(\mathbf{r},t) \hat{\mathbf{E}}(\mathbf{r},t)
\psi_\eta(\mathbf{r},t).
\end{align}

%
%We consider here a general case including the BEC of dipole Bose particles, ${\bf d}_{00}\neq 0$.

For the particles with $\eta \neq 0$, one can use the plain-wave ansatz: $\psi_{\eta\neq 0}(\mathbf{r},t)=\sum_{\mathbf{p}}c_{\eta\mathbf{p}}(t)\exp(\text{i}\mathbf{p}\mathbf{r})$, where $c_{\eta\mathbf{p}}(t)=c_{\eta\mathbf{p}}(0)\exp(-\text{i}E_\eta(\mathbf{p})t)$ with the corresponding energy $E_\eta(\mathbf{p})=p^2/2m+\Delta_\eta$. 
For the condensate, $\eta=0$,  the Bogoliubov transformation reads  $\psi_{0}(\mathbf{r},t)=\sum_{\mathbf{p}}\left[\sqrt{n_c}\delta(\mathbf{p}) +u_\mathbf{p}b_\mathbf{p}(t) +v_\mathbf{p}b_{-\mathbf{p}}^\dagger(t)\right]\exp(\text{i}\mathbf{p}\mathbf{r})$,  where $n_c$ is the density of the condensate, $u_{\textbf{p}}$ and $v_{\textbf{p}}$ are the Bogoliubov transformation coefficients, and $b_\mathbf{p}(t)=b_\mathbf{p}(0)\exp(- \text{i}\varepsilon_\mathbf{p}t)$ is the annihilation operator for Bogoliubov quasi-particle (bogolon).
The spectrum of bogolons is given by~\cite{LANDAU198079} $\varepsilon_\mathbf{p}=sp\sqrt{1+p^2\xi^2}$ with $s=\sqrt{n_c g/m}$ being the sound velocity and $\xi=(2ms)^{-1}$ being the healing length.
Then, the first term in~\eqref{EqTerms} reads:

\begin{align}
\label{EqV1}
\hat{V}_1&=
-\mathbf{d}_{00}
\int
d\mathbf{r}\psi_{0}^\dagger(\mathbf{r},t)
\hat{\mathbf{E}}(\mathbf{r},t)
\psi_0(\mathbf{r},t) \\
\nonumber
&=-\mathbf{d}_{00}
\sum_{\mathbf{p}^\prime \mathbf{p}}
\left[
\sqrt{n_c}\delta(\mathbf{p}^\prime)
+u_{\mathbf{p}^\prime}b^\dagger_{\mathbf{p}^\prime}(t)
+v_{\mathbf{p}^\prime}b_{-\mathbf{p}^\prime}(t)
\right]\\
\nonumber
&~~\times
\int d\mathbf{r}\exp(-\text{i}\mathbf{p}^\prime\mathbf{r})
\hat{\mathbf{E}}(\mathbf{r},t)
\exp(\text{i}\mathbf{p}\mathbf{r})
\left[
\sqrt{n_c}\delta(\mathbf{p})
+u_{\mathbf{p}}b_{\mathbf{p}}(t)
+v_{\mathbf{p}}b_{-\mathbf{p}}^\dagger(t)
\right].
\\
\nonumber
 \end{align}
In what follows, let us only focus on the light absorption processes, thus disregarding the terms containing $\hat{\mathbf{E}}^\dagger_0$.
Moreover, at low temperatures only the processes accompanied by the emission (not absorption) of bogolons are considerable, thus only the terms containing $b^\dagger$ and $b^\dagger b^\dagger$ matter. 
After the integration, 
Eq.~\eqref{EqV1} reads:

\begin{align}
\label{EqV1k}
\hat{V}_1 = & - (2 \pi)^d \sum_{\mathbf{p}^\prime \mathbf{p}}\delta (\mathbf{p'-p-k})  \left [  \sqrt{n_c} \left ( \delta(\mathbf{p})u_{p'}b^{\dagger}_{p'}(t) + \delta(\mathbf{p'})v_{p}b^{\dagger}_{-p}(t) \right ) \right. \\\nonumber
        &~~~~~~~ \left.  + u_{p'}v_{p} b^{\dagger}_{p'}(t) b^{\dagger}_{-p}(t) \right ]  \mathbf{d}_{00} \cdot \hat{\mathbf{E}}_0 (t),\nonumber
\end{align}
where the trivial case ($\mathbf{k} = 0$) was disregarded.
Thus, $\hat{V_1}$ describes a possible direct light absorption by the BEC with a possible excitation of single or two bogolons in the BEC. It should be noted that these processes are possible only in the BEC of atoms having a nonzero dipole moment in its ground state, ${\bf d}_{00}\neq 0$.

Following a similar procedure gives other interaction terms,
\begin{align}
 \hat{V}_2 & =  - (2 \pi)^d \sum_{\mathbf{p}^\prime \mathbf{p}}
 \sum_{\eta}
 \delta(\mathbf{p'-p-k}) \times u_{p'}b^{\dagger}_{p'}(t) c_{\eta \mathbf{p}}(t)  \mathbf{d}_{0\eta} \cdot \hat{\mathbf{E}}_0 (t)\\
\hat{V}_3 & =  - (2 \pi)^d \sum_{\mathbf{p}^\prime \mathbf{p}}
\sum_{\eta}
\delta(\mathbf{p'-p-k})   \label{eq:v3} \times c_{\eta \mathbf{p'}}^{\dagger}(t) \left [ \sqrt{n_c} \delta(\mathbf{p})+ v_p b^{\dagger}_{-p}(t) \right ] \mathbf{d}_{\eta 0} \cdot \hat{\mathbf{E}}_0 (t) 
\\
\hat{V}_4 &= -\left(2\pi \right)^d \sum_{\mathbf{p}^\prime \mathbf{p}}
\sum_{\eta\eta'}
\delta(\mathbf{p'-p-k}) \times c_{\eta^\prime \mathbf{p}^\prime}^\dagger \left(t\right) c_{\eta \mathbf{p}}\left(t\right) \mathbf{d}_{\eta^\prime \eta} \cdot \hat{\mathbf{E}}_0 \left(t\right).
\end{align}

Before the system being irradiated, all the particles are in BEC.
Therefore, $\hat{V_2}$ and $\hat{V_4}$ terms can be disregarded in later consideration.

Let us now discuss the physical meaning of the processes incorporated in $V_3$ term. 
The first term in $V_3$, which is $\propto \sqrt{n_c}$, describes the absorption of the photon with energy $\omega_{\bf k}$ and momentum $\bf k$ accompanied by the direct excitation of the atom from the BEC state $\eta=0;\,\,{\bf p}=0$ to the noncondensed state with $\eta'\neq 0$ and momentum ${\bf p}'={\bf k}$ with the energy $E_{\eta'}({\bf p}'={\bf k})=\omega_{\bf k}$.

The second term in $V_3$ describes the transition of the BEC atom to the final state with $\eta'\neq 0$ and momentum ${\bf p}'={\bf p}+{\bf k}$ with an arbitrary value of ${\bf p}$. 
To conserve the total momentum, a single bogolon with energy $\varepsilon_{-{\bf p}}$ is excited in the BEC carrying away the missing momentum $-{\bf p}$ in such a way that the total momentum in the system is conserved.
Below, the processes described by $V_1$ and $V_3$ terms are analyzed in more detail.

\section{Absorption probability}
According to the Fermi golden rule, the absorption probability~\cite{chuang2012physics} for different interaction channels reads as 

\begin{equation}
    \alpha = \frac{2\pi}{\hbar} |\bra{f} \hat{V} \ket{i}|^2 \delta\left(E_f + E_i - \omega\right),
\end{equation}
where $\ket{i}$ is the initial (unperturbed) state. 
The perturbation, which is the EM field, results in the transitions from the initial state to the final state $\ket{f}$.
%
%Let us consider the term $V_1$ in Eq.~\eqref{EqV1k} and write the light absorption probability including the single (1b) and two-bogolon (2b) emission processes, respectively.
%
The $\hat{V}_1$ part of BEC-light interaction Hamiltonian contains two different processes. 
In the first term in~\eqref{EqV1k}, the initial state of the system is the unperturbed BEC and the photon with energy $\omega_{\bf k}$, whereas the final state corresponds to the presence of single bogolon in the system with energy $\varepsilon_{\bf k}$. 
Thus, the corresponding probability can be written as
%$\\\\$
%For 1b,
%
%
%
\begin{equation}
\label{alpha1}
\alpha_{1b}^1=
\frac{2\pi}{\hbar}
|\mathbf{d}_{00}\cdot \hat{\mathbf{E}}_0|^2
|u_{-\mathbf{k}}+v_{\mathbf{k}}|^2
\delta(\varepsilon_{\mathbf{k}}-\omega_{\mathbf{k}}).
\end{equation}
Here, 1b stands for one-bogolon--mediated processes, the $\delta$-function indicates the energy conservation law for the direct process, 
photon transforming its energy into a bogolon. 

The probability due to the second term in~\eqref{EqV1k} describes the transitions from the BEC accompanied by two bogolons (2b),
\begin{align}
\label{alpha12b}
\alpha_{2b}^1&=
\frac{2\pi}{\hbar}
|\mathbf{d}_{00}\cdot \hat{\mathbf{E}}_0|^2
\sum_\mathbf{p}
|u_{\mathbf{p}+\mathbf{k}}v_{\mathbf{p}}|^2
\delta(\varepsilon_{\mathbf{p}+\mathbf{k}}+\varepsilon_{-\mathbf{p}} -\omega_{\mathbf{k}}).
\end{align}
Expressions~\eqref{alpha1} and~\eqref{alpha12b} describe light absorption with a direct transfer of the photon energy to the excitations of the BEC. 
Since the phase velocity of bogolons is much lower than the speed of light, $\varepsilon_{\bf k}\neq\omega_{\bf k}$ for any finite values of $\bf k$. 
Thus, Eq.~\eqref{alpha1} does not give contribution.
In contrast, the probability~\eqref{alpha12b} can be finite. 
Its value for typical parameters characterizing atomic BECs of different dimensionalities is analyzed below.

A similar analysis can be applied to  expression~\eqref{eq:v3}. 
The first term in $\hat{V}_3$ describes the direct transition of an individual atom energy 
from the ground to an excited state under photon absorption. 
This interaction channel does not involve any bogolons,
\begin{equation}
\label{alpha30b}
\alpha_{0b}^3 = \frac{2\pi}{\hbar} n_c \sum_{\eta\neq 0}   | \mathbf{d}_{\eta 0} \cdot \hat{\mathbf{E}}_0 |^2    
 \delta ( E_{\eta}(\mathbf{k}) -\omega_{\mathbf{k}}  ).     
\end{equation} 
The second term in \eqref{eq:v3} involves 1b  processes, and the corresponding probability reads
\begin{equation}
\label{alpha31b}
\alpha_{1b}^3 = \frac{2\pi}{\hbar}  \sum_{\eta,\mathbf{p}}   | \mathbf{d}_{\eta 0} \cdot \hat{\mathbf{E}}_0 |^2
 |v_{\mathbf{p}}|^2  \delta ( E_{\eta}(\mathbf{p+k}) + \varepsilon_{\mathbf{-p}} - \omega_{\mathbf{k}} ).
\end{equation}
Thus, in zero-temperature limit, photons can only be absorbed by the BEC in three different channels:
(i) by generating an excited boson; (ii) by generating a single bogolon and an excited boson; (iii) by generating a pair of bogolons from the condensate.

In the most interesting case $p\xi\ll1$ when the dispersion of the Bogoliubov excitations of the BEC can be approximated by the linear dispersion as $\varepsilon_p \approx sp$, analytical results for the probabilities of these processes can be  found. 
Under this approximation, the Bogoliubov transformation coefficients can be written as
\begin{equation}
\label{trans_coeffi}
u_{\mathbf{p}} \approx \sqrt{\frac{ms}{2| \mathbf{p}|}} ,\quad v_{\mathbf{p}} \approx - \sqrt{\frac{ms}{2| \mathbf{p}|}}. 
\end{equation}
Furthermore, we analytically calculate  the absorption probabilities given by 
Eqs.~\eqref{alpha12b},~\eqref{alpha30b} and~\eqref{alpha31b} in various dimensions.

\subsection{Absorption probability \texorpdfstring{$\alpha_{2b}^1$}{Lg}}

\noindent
Let us, first, find $\alpha_{2b}^1$ in different dimensionalities by substituting Eq.~\eqref{trans_coeffi} in Eq.~\eqref{alpha12b}, yielding
\begin{align}
\alpha_{2b}^1 &=  \frac{\pi m^2s}{2\hbar}|\mathbf{d}_{00}\cdot \hat{\mathbf{E}}_0|^2 \frac{1}{(2\pi\hbar)^d} \int d\mathbf{p} \frac{1}{|\mathbf{p}||\mathbf{p+k}|} \delta \left( |\mathbf{p+k}|+|\mathbf{p}| - \frac{\omega_{\mathbf{k}}}{s} \right ). 
\end{align}
% For a non-zero $\mathbf{k}$, the integral is well-behaved (does not diverge).

In case of 1D BEC, the absorption probability is 
\begin{eqnarray}
\label{eq:alpha_2b_1d}
a^1_{2b,\textrm{1D}}&=&
\frac{m^2s}{4 \hbar^2}\abs{\textbf{d}_{00}\cdot\hat{\textbf{E}_0}}^2
I_{2b,1D}^1
=
\frac{2m^2s}{ \hbar^2}\abs{\textbf{d}_{00}\cdot\hat{\textbf{E}_0}}^2\frac{\Theta\left[\omega_{\mathbf{k}}/s-k_x\right]}{(\omega_{\mathbf{k}}/s)^2-k_x^2},
\end{eqnarray}
where 
\begin{equation}
\label{eq:I12b1d}
I^1_{2b,\textrm{1D}}
=\int dp\frac{1}{\abs{p}\abs{p+k_x}}\delta\left(\abs{p+k_x}+\abs{p}-\frac{\omega_{\mathbf{k}}}{s}\right).
\end{equation}
In Eq.~\eqref{eq:I12b1d}, the momentum $k\equiv k_\parallel$ inherent from the photon is just the component which is parallel to the 1D sample.
It should be noted, that the frequency of light $\omega_{\bf k}$ depends on both the parallel and vertical components of $\mathbf{k}\left(k_\parallel, k_\perp\right)$.

In the 2D case, the absorption probability reads (\textcolor{blue}{See the details in Appendix}),
\begin{align}
\alpha_{2b,\textrm{2D}}^1 &=  
\frac{sm^2}{8\pi \hbar^3}
|\mathbf{d}_{00}\cdot \hat{\mathbf{E}}_0|^2
I_{2b,2D}^1 =
\frac{sm^2}{4\hbar^3}
|\mathbf{d}_{00}\cdot \hat{\mathbf{E}}_0|^2
\frac{\Theta[\omega_{\mathbf{k}}/s - k_\parallel]}{\sqrt{(\omega_{\mathbf{k}}/s)^2-k_\parallel^2}}.
\end{align}
In 3D, we have
\begin{align}
\alpha_{2b,\textrm{3D}}^1 &=  
\frac{sm^2}{16\pi^2 \hbar^4}
|\mathbf{d}_{00}\cdot \hat{\mathbf{E}}_0|^2
I_{2b,3D}^1 =
\frac{sm^2}{8\pi \hbar^4}
|\mathbf{d}_{00}\cdot \hat{\mathbf{E}}_0|^2
\Theta[\omega_{\mathbf{k}}/s-k].
\end{align}

\subsection{Absorption probability \texorpdfstring{$\alpha_{1b}^3$}{Lg}}

By plugging Eq.~\eqref{trans_coeffi} into Eq.~\eqref{alpha31b}, we attain 
\begin{align}
\label{approx}
\alpha_{1b}^3  &= \frac{m\pi}{\hbar}\frac{s}{(2\pi\hbar)^d}
\sum_{\eta}|\textbf{d}_{\eta0}\cdot\hat{\textbf{E}_0}|^2
\int d\textbf{p}\frac{1}{p}\delta(\Delta_\eta+sp-\omega_\mathbf{k}).
\end{align}
Here, several approximations have been assumed. First, the bogolon dispersion is linear, $\varepsilon_{\bf p}\approx sp$. 
This approximation implies the dropping of the $p^2$ and other higher-order terms in the bogolon dispersion. 
At the same level of approximation, we must disregard the kinetic energy of individual atom, $({\bf p}+{\bf k})^2/2m$, assuming $E_\eta({\bf p}+{\bf k})\approx\Delta_\eta$ in  Eq.~\eqref{alpha31b}. 
%Thus, after all these appoximations, we arrive at Eq.\eqref{approx}.}  

This formula, Eq.~\eqref{approx}, produces different results for BECs of different dimensions
(\textcolor{blue}{See the details in Appendix}),

\begin{align}
\alpha_{1b,\textrm{1D}}^3  
&=
\frac{m}{2\hbar^2}\sum_{\eta}
|\textbf{d}_{\eta0}\cdot\hat{\textbf{E}_0}|^2
I_{1b,1D}^3  
=
\frac{ms}{\hbar^2}\sum_{\eta}|\textbf{d}_{\eta0}\cdot\hat{\textbf{E}_0}|^2
\frac{\Theta[\omega_{\mathbf{k}}-\Delta_\eta]}{\omega_{\mathbf{k}} - \Delta_\eta},
\\
\alpha_{1b,\textrm{2D}}^3   &= 
\frac{m}{4\pi\hbar^3}\sum_{\eta}
|\textbf{d}_{\eta0}\cdot\hat{\textbf{E}_0}|^2
I_{1b,2D}^3  
=
\frac{m}{2\hbar^3}\sum_{\eta}
|\textbf{d}_{\eta0}\cdot\hat{\textbf{E}_0}|^2
\Theta[\omega_{\mathbf{k}}-\Delta_\eta],
\\
\alpha_{1b,\textrm{3D}}^3  &= 
\frac{m}{8\pi^2\hbar^4}\sum_{\eta}
|\textbf{d}_{\eta0}\cdot\hat{\textbf{E}_0}|^2
I_{1b,3D}^3  =
\frac{m}{2\pi s\hbar^4}\sum_{\eta}\abs{\textbf{d}_{\eta0}\cdot\hat{\textbf{E}_0}}^2(\omega_\mathbf{k}-\Delta_\eta)\Theta[\omega_\mathbf{k}-\Delta_\eta].
\end{align}

Let us compare the absorption probability for the 1b and 2b processes.

\begin{figure}[ht]
\includegraphics[scale=0.4]{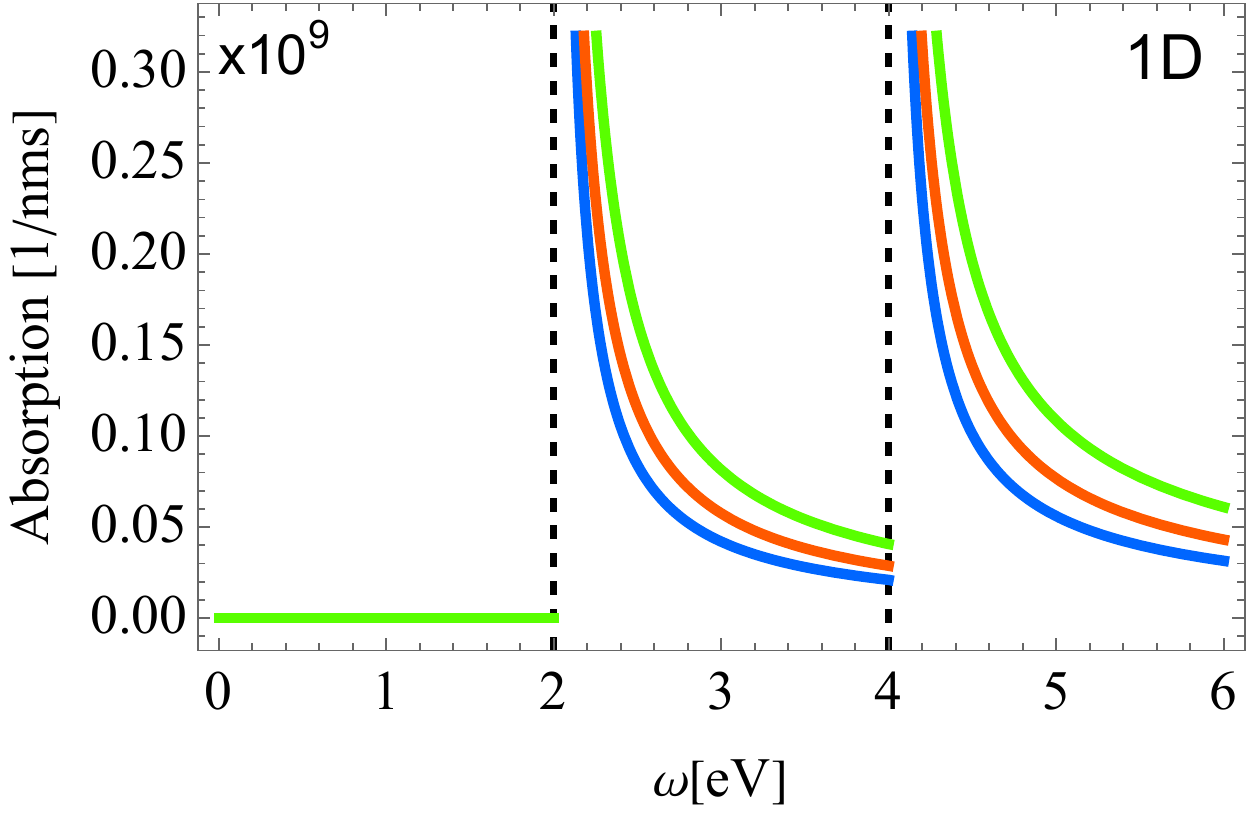}
\includegraphics[scale=0.4]{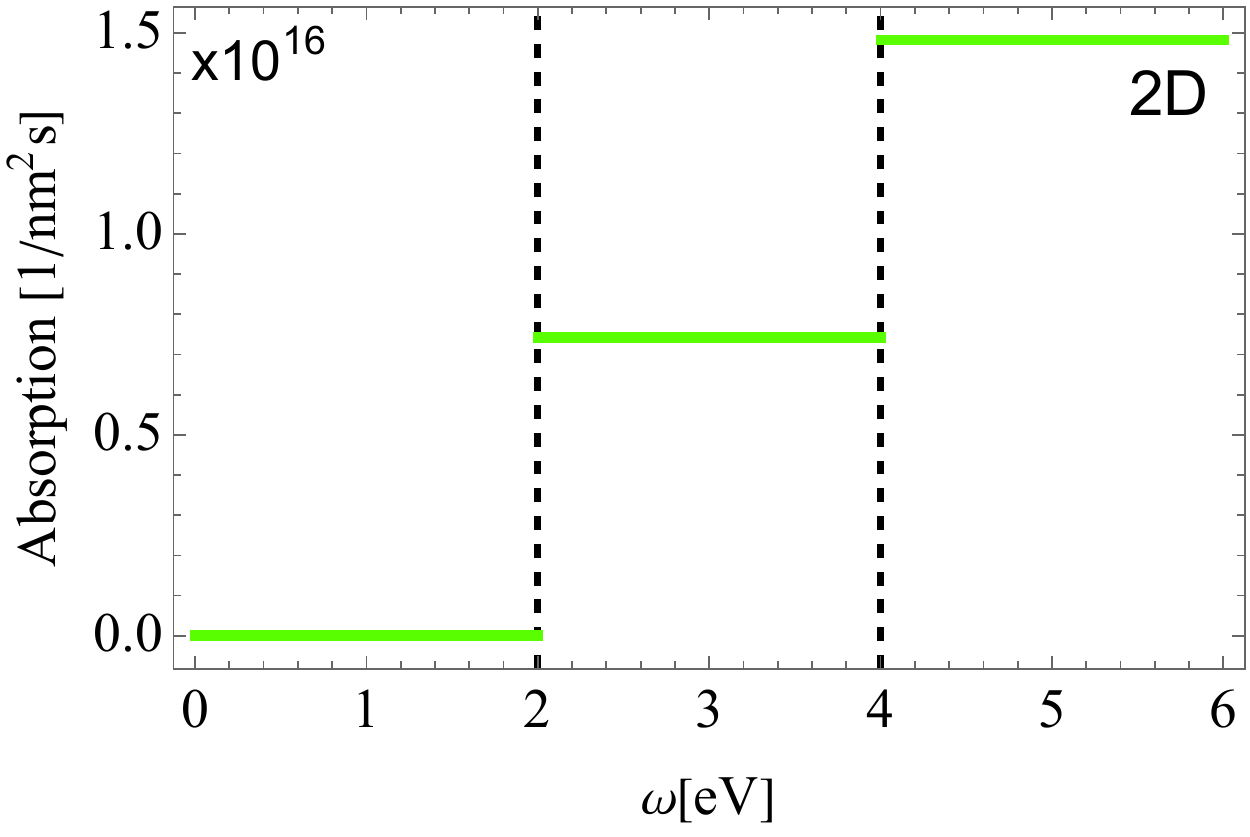}
\includegraphics[scale=0.4]{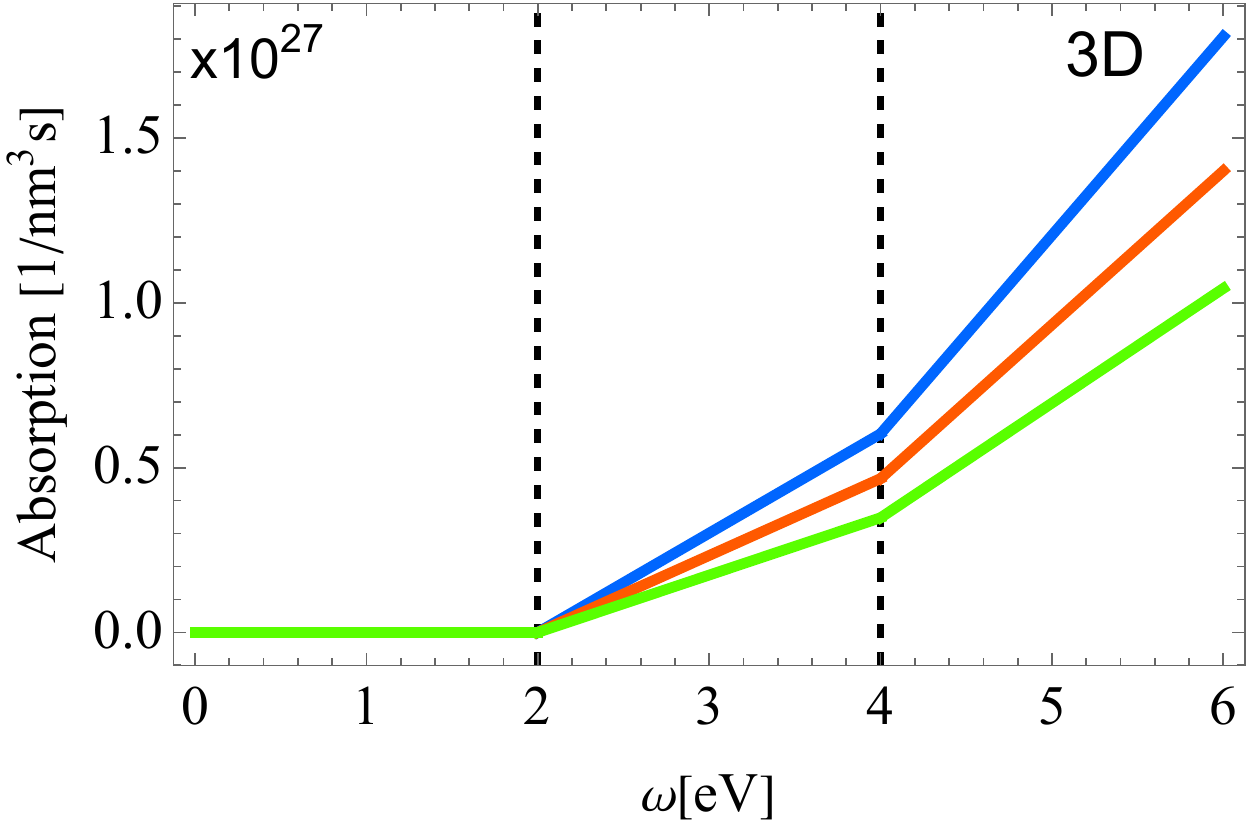}
\caption
{
Spectrum of the absorption probability in each dimensionality for 1b processes. The parameters used are: $m=1.443\times10^{-25}$~kg is the mass of atom. $\Delta_\eta= \eta\cdot 2$eV is the energy spectrum of the internal motion, $|\mathbf{d}_{\eta0}\cdot\hat{\mathbf{E}}_0|=2\times10^{-3}$eV is the transition matrix element with $\textbf{d}_{\eta 0}=3.584\times10^{-29}C\cdot m$ the transition dipole matrix element; $\hat{\mathbf{E}}_0=100$ KV/cm is electric field intensity. $k=\omega/c$ is the wavevector of  light. 
For 1D, the condensate densities are $n_{1d}=80~\mu $m$^{-1}$(blue), $n_{1d}=150~\mu $m$^{-1}$(red), $n_{1d}=300~\mu $m$^{-1}$(green). For 2D, $n_{2d}=100~\mu $m$^{-2}$. 
For 3D, the condensate densities are $n_{3d}=250~\mu $m$^{-3}$(blue), $n_{3d}=320~\mu $m$^{-3}$(red), $n_{3d}=400~\mu $m$^{-3}$(green).}
\label{FIG1}
\end{figure}

\begin{figure}[ht]
\includegraphics[scale=0.4]{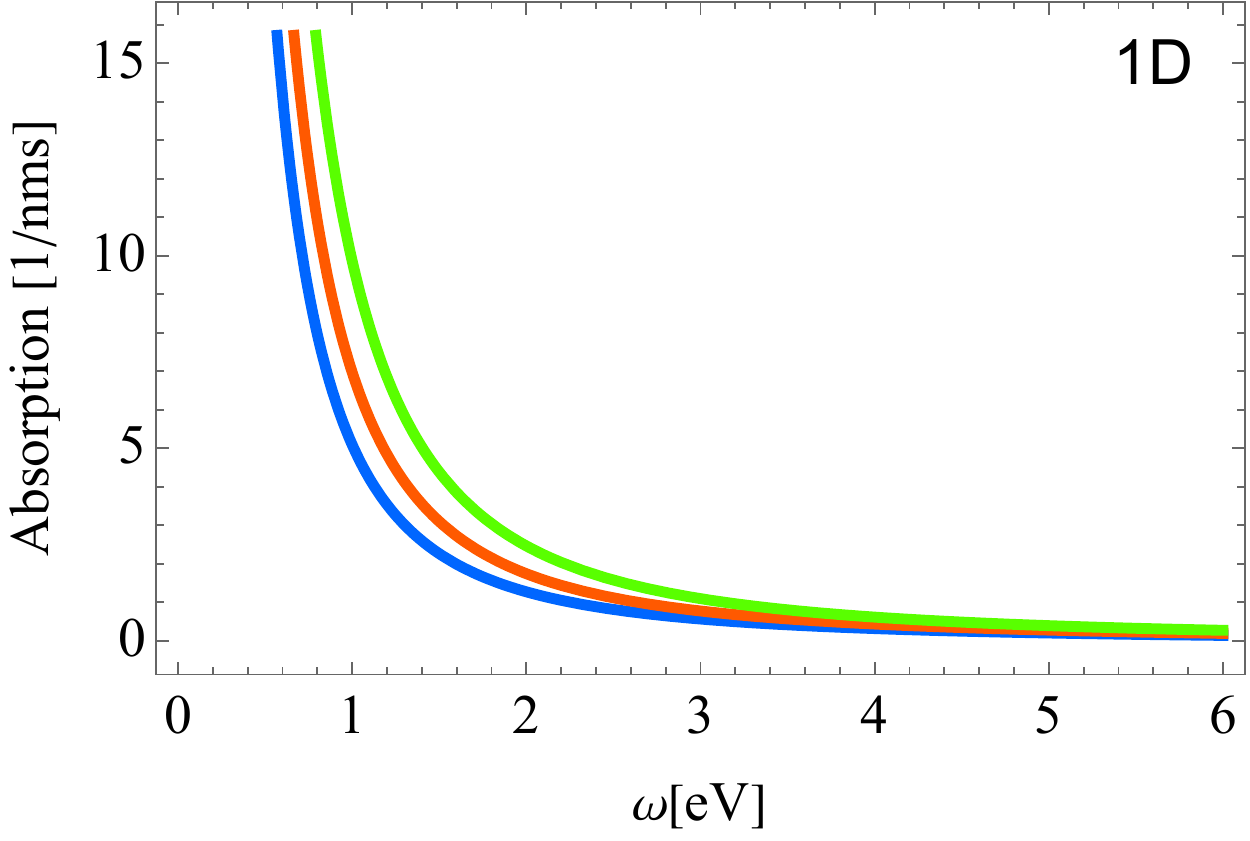}
\includegraphics[scale=0.4]{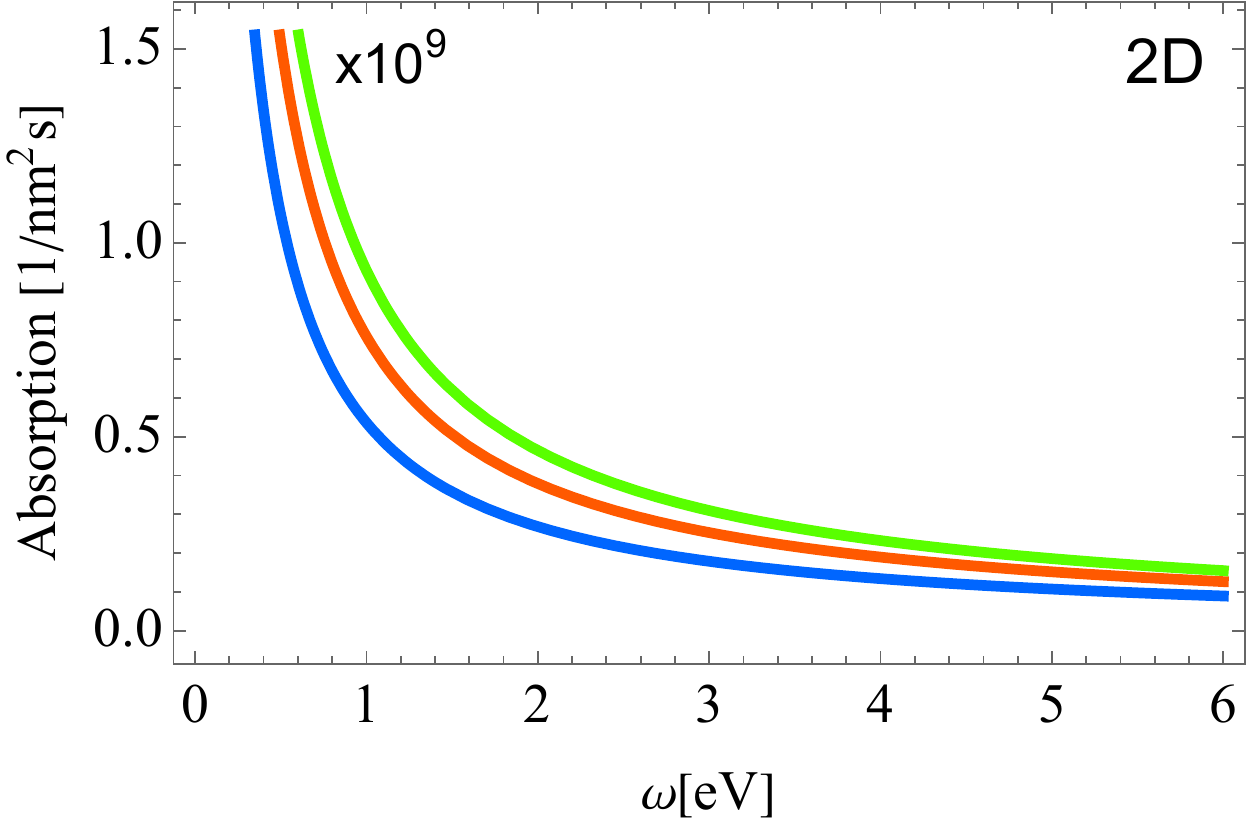}
\includegraphics[scale=0.4]{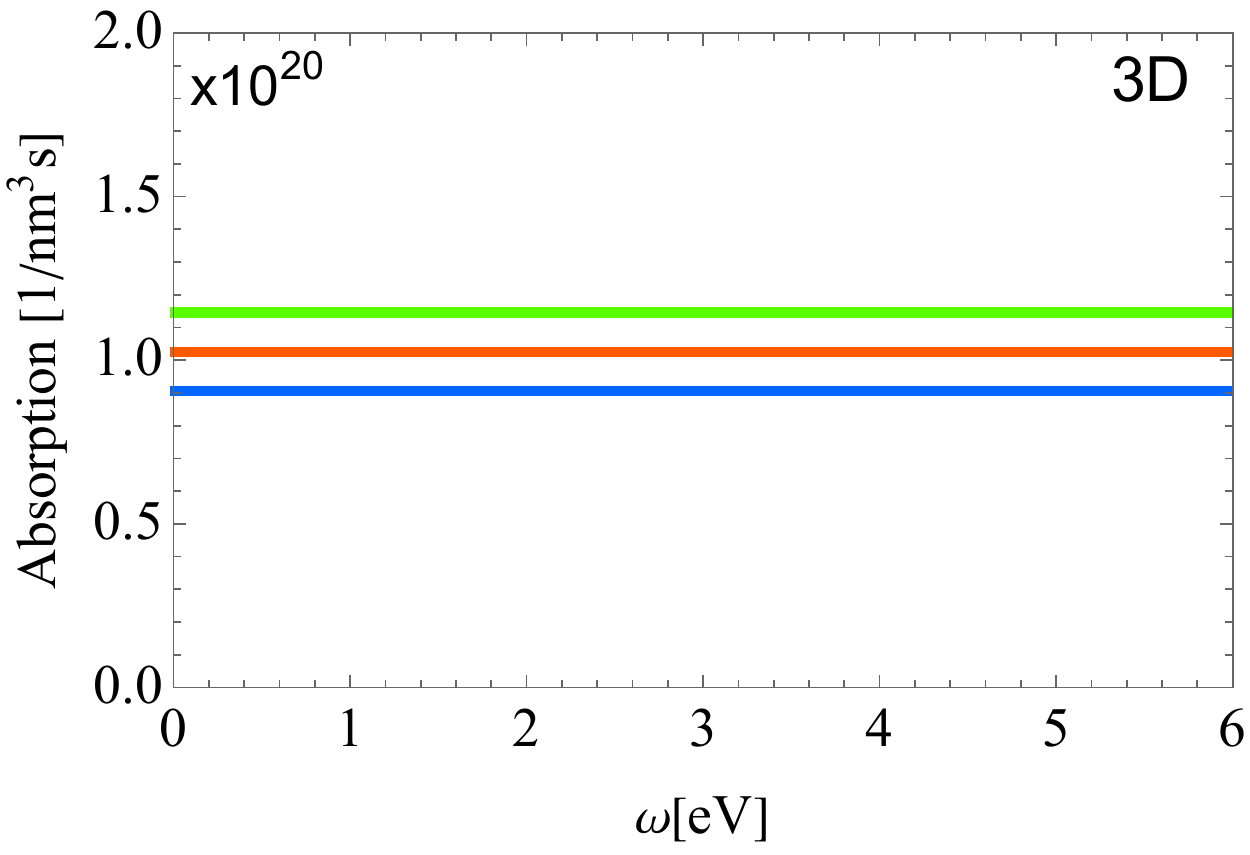}
\caption
{
Absorption probabilities for 2b processes in different dimensionalities with various condensate densities. For 2D, the condensate densities are 
$n_{2d}=100~\mu$m$^{-2}$(blue), $n_{2d}=200~\mu$m$^{-2}$(red), $n_{2d}=300~\mu$m$^{-2}$(green).
The other parameters are taken the same as in Fig.~\ref{FIG1}.
}
\label{FIG2}
\end{figure}

\section{Discussion}
As concerns expression~\eqref{alpha30b},
it produces a series of resonances corresponding to direct transitions of the atoms from BEC without  the excitation of the condensate density. 
Despite the absence of bogolons under this photon absorption processes, its magnitude is determined by the BEC density $n_c$.

Figure~\ref{FIG1} shows the absorption probabilities in case of 1b processes for different dimensionalities of the system and for various condensate densities. 
Below certain threshold, $\Delta=$2~eV in our case, the absorption in all the 1D, 2D and 3D samples is absent. 
At higher frequencies $\omega>\Delta$, in 1D case, the absorption experiences a comb-like behavior with the peaks at $\omega_{\bf k}=\eta\Delta$. %\textcolor{blue}{Where we find this expression from?}
By increasing the condensate density, the absorption probability also increases. 
In the 2D case, the absorption reveals a step-like behavior, and it does not depends on the condensate density. 
In 3D, the absorption probability curve reminds a broken threshold.
Interesting to note, that it decreases with the increase of the condensate density, which is opposie to the 1D case. 

Figure~\ref{FIG2} shows the absorption probabilities for 2b processes and for different condensate densities. 
By increasing the condensate density, the absorption probability increases in all dimensionalities. 
For 1D and 2D, we observe an exponential decay, and in 3D, the absorption probability is a constant. 

There are several differences between 1b and 2b process.
In 2D, 2b processes-mediated absorption does depend on the condensate density as compared to 1b processes, which do not experience any dependence on the condensate density. 
Another principle difference between 1b and 2b processes is that for 2b process, the absorption is finite even at small frequencies ($\omega<\Delta$). 
Moreover, it is more pronounced at smaller frequencies.

The theory developed above is based on the model, which disregards the disorder and interaction between photo-excited particles and BEC. 
These effects result in finite lifetime of the particles, changing the density of state in the system, which, in order, might modify the frequency dependence of the light
absorption coefficient, 
which is defined as the ratio of number of absorbed photons and number of incident photons, $\alpha(\hbar \omega) = \hbar \omega W/P~\cite{chuang2012physics}$,
where $W$ is absorption probability and $P$ is the average of the Poynting flux for the light intensity, $P=c\epsilon_0E_0^2/2$, 
where $\epsilon_0$ is the vacuum permittivity. 

%This important questions are beyond of the present paper consideration and may be the subject of the further theory developing. We also ignoring the possible...
In the case of a disordered BEC or finite lifetime of bogolons, the $\delta$-functions in the equations for the absorption probability terms has to be widened into the Lorenz form. 
The microscopic analysis of the disorder or particle collisions is a separate question, which we leave beyond the scope of the present paper. 

The other specific feature of the presented theory is that we consider BEC in different dimensions to be spatially uniform, such that $n_c=const$. 
In case of 0D trapped or spatially-modulated BECs the condensate density in equilibrium becomes spatially-dependent, $n_c({\bf r})$. 
Thus, the system preserves the translation invariance.
In that case, the Bose  particles momenta (and bogolons' ones) ${\bf p}$ are not conserved violating the momentum conservation law under optical transitions. 
A careful analysis of light absorption in these BECs require a separate consideration.

\section{Conclusions}

We studied the response of a cold atomic gas in the BEC phase to an external electromagnetic field by calculating the absorption probabilities. 
For that, we used the standard Bogoliubov theory, extending it to the case of Bose particles possessing internal degrees of freedom.
We show, that several specific processes might occur if the atomic gas is in the BEC state.
In particular, two types of transitions occur, which contribute to light absorption by the system. 
The processes of the first kind involve an excitation of an atom accompanied by an emission of a bogolon -- the quantum of BEC density fluctuations. 
%We show, that this type of processes are dicteed by the strong conservation laws of momentum and enerfy udder the optical transitons in the BEC. 
The processes of the second type involve creating two bogolons with different momenta. We demonstrate that the one-bogolon processes are dominant in a broad range of the external EM field frequencies except for the small-frequency region.
Moreover, the light absorption in different dimensionalities depends differently on the condensate density.
For one-bogolon processes, the absorption increases with larger condensate densities in 1D. However, in 3D, by increasing the condensate density, the absorption decreases. 
In 2D, the absorption does not depend on the condensate density. 
For the process involving pairs of bogolons, in all the dimensions, the absorption increases with the increases of condensate density.

\ack
%\section{ACKNOWLEDGMENTS}
We acknowledge the support by the Institute for Basic Science in Korea (Project No.~IBS-
R024-D1) and by the Ministry of Science and Higher Education of the Russian Federation (Project No.~075-15-2020-797 (13.1902.21.0024)). 
M.S. was partially supported by Startup Fund for Doctoral Research from Beijing University of Technology.

\section*{Appendix}
This appendix provides the details of evaluation of all the integrals for the absorption probabilities in various dimensionalities. 
First, to find $I_{2b,\textrm{2D}}^1$ we set $q=|\mathbf{p}+{\bf k}_{||}|$. 
Thus, the in-plain angle $\theta$ between momenta $\bf p$ and ${\bf k}_{||}$ reads
\begin{gather}
\sin \theta = \frac{\sqrt{[(p+k_{||})^2-q^2][q^2-(p-k_{||})^2]}}{2pk_{||}}, 
\end{gather}
and 
\begin{gather}
\label{theta}
d\theta = - qdq/(pk_{||}\sin\theta) =-2\frac{qdq}{\sqrt{[(p+k_{||})^2-q^2][q^2-(p-k_{||})^2]}}.
\end{gather}
Substitution Eq.~\eqref{theta} in $I_{2b,\textrm{2D}}^1$ yields

\begin{eqnarray}
\label{substitutt}
&&I_{2b,\textrm{2D}}^1 = \int\limits_{0}^{\infty} p dp \int\limits_{0}^{2\pi} d\theta \frac{1}{p|\mathbf{p}+{\bf k}_{||}|} \delta( |\mathbf{p}+{\bf k}_{||}|+p - \omega_{\mathbf{k}}/s ) \nonumber \\
  &&= 4 \int\limits_{0}^{\infty} dp \int\limits_{|p-k_{||}|}^{|p+k_{||}|} dq 
\frac{ \delta \left ( q+p - \omega_{\mathbf{k}}/s \right )}{\sqrt{[(p+k_{||})^2-q^2][q^2-(p-k_{||})^2]}}.
\end{eqnarray}

%\noindent
%
%
%
%
Furthermore, we employ the Lagrange multipliers by the change of variables, $q+p = x$ and $q-p = y$, thus
\begin{gather}
\label{Lagrenge}
\int\limits_{0}^{\infty} dp \int\limits_{|p-k_{||}|}^{|p+k_{||}|} dq = \int\limits_{k_{||}}^{\infty} dx \int\limits_{-k_{||}}^{k_{||}} dy \left | \frac{\partial(p,q)}{\partial(x,y)} \right |.
\end{gather}
Substituting Eq.~\eqref{Lagrenge} in Eq.~\eqref{substitutt} gives
\begin{gather}
I_{2b,\textrm{2D}}^1 = 2 \int\limits_{k}^{\infty} dx \int\limits_{-k_{||}}^{k_{||}} dy \frac{ \delta \left (x - \omega_{\mathbf{k}}/s \right ) }{ \sqrt{(x^2-k_{||}^2)(k_{||}^2-y^2)} }  \\ \nonumber
  = 2 \pi  \int\limits_{k_{||}}^{\infty} dx \frac{ \delta \left (x - \omega_{\mathbf{k}}/s \right ) }{ \sqrt{x^2-k_{||}^2} } 
  =  \frac{ 2\pi\Theta (\omega_{\mathbf{k}}/s-k_{||}) }{\sqrt{(\omega_{\mathbf{k}}/s)^2 -k_{||}^2}}. 
\end{gather}
%
%
%
%\noindent

A similar method can be employed to calculate $I_{2b,\textrm{3D}}^1$, 
\begin{gather}
I_{2b,\textrm{3D}}^1 = 2\pi \int\limits_{0}^{\infty} p^2 dp \int\limits_{|p+k|}^{|p-k|} \frac{qdq}{pk} \frac{1}{pq} \delta \left ( q+p - \omega_\mathbf{k}/s \right ) \nonumber \\
=\frac{\pi}{k} \int\limits_{k}^{\infty} dx \int\limits_{-k}^{k} dy \delta \left (x - \omega_\mathbf{k}/s \right )
= 2 \pi \Theta (\omega_\mathbf{k}/s-k).
\end{gather}

Next, for 1b processes
\begin{gather}
I_{1b,\textrm{1D}}^3 =
\int d\textbf{p}\frac{1}{|p|}\delta\left(\frac{\Delta_\eta}{s}+p-\frac{\omega_{\mathbf{k}}}{s}
\right)
=
\int\limits^{\infty}_{-\infty}\frac{dp}{|p|}\delta\left(\frac{\Delta_\eta}{s}+p-\frac{\omega_{\mathbf{k}}}{s}\right)
=
2s\frac{\Theta[\omega_{\mathbf{k}}-\Delta_\eta]}{\omega_{\mathbf{k}} - \Delta_\eta},
\end{gather}
in the case of 1D BEC. 
The corresponding expression for 2D condensate reads
\begin{gather}
I_{1b,\textrm{2D}}^3=
\int d\textbf{p}\frac{1}{p}\delta\left(\frac{\Delta_\eta}{s}+p-\frac{\omega_{\mathbf{k}}}{s}\right)
=
\int\limits^{\infty}_0dp\int\limits^{2\pi}_0d\theta\delta\left(\frac{\Delta_\eta}{s}+p-\frac{\omega_{\mathbf{k}}}{s}\right)
\\
\nonumber
=
2\pi\int\limits^{\infty}_0dp\delta\left(\frac{\Delta_\eta}{s}+p-\frac{\omega_{\mathbf{k}}}{s}\right)
=
2\pi\Theta[\omega_{\mathbf{k}}-\Delta_\eta],
\end{gather}
and, finally, for 3D BEC one finds
\begin{gather}
I_{1b,\textrm{3D}}^3=
\int d\textbf{p}\frac{1}{p}\delta\left(\frac{\Delta_\eta}{s}+p-\frac{\omega_{\mathbf{k}}}{s}\right)
=
\int\limits^{\infty}_0pdp\int\limits^{\pi}_0\sin{\theta}d\theta\int\limits^{2\pi}_0d\phi\delta\left(\frac{\Delta_\eta}{s}+p-\frac{\omega_{\mathbf{k}}}{s}\right)
\\
=
\nonumber
4\pi
\int\limits^{\infty}_0pdp\delta\left(\frac{\Delta_\eta}{s}+p-\frac{\omega_\mathbf{k}}{s}\right)=
\frac{4\pi}{s}(\omega_\mathbf{k}-\Delta_\eta)\Theta[\omega_\mathbf{k}-\Delta_\eta].
\end{gather}
These expressions are analyzed in the main text.

\section*{References}
\bibliographystyle{iopart-num}
\bibliography{main}

\end{document}